\documentclass[prx,superscriptaddress,twocolumn,longbibliography]{revtex4-1}

\usepackage{amsmath}
\usepackage{amssymb}
\usepackage{amsxtra,color}
\usepackage{latexsym}
\usepackage{graphicx}
\usepackage[utf8]{inputenc}
\usepackage{MnSymbol}	% use for sumint

\usepackage{hyperref}

\definecolor{MyDarkGreen}{rgb}{0,0.6,0}
\definecolor{MyDarkBlue}{rgb}{0,0,0.8}
\definecolor{MyDarkRed}{rgb}{0.6,0,0.3}

\hypersetup{breaklinks=true, colorlinks=true,plainpages=true, linktocpage=true, linkcolor=MyDarkBlue, citecolor=MyDarkGreen, urlcolor=MyDarkRed, pdfborder={0 0 0},%
pdfauthor={},%
pdfsubject={Research article},
pdftitle={}%
}

\newcommand{\ba}{\begin{align}}
\newcommand{\ea}{\end{align}}
\newcommand{\be}{\begin{equation}}
\newcommand{\ee}{\end{equation}}
\newcommand{\bra}[1]{\langle  #1 |}
\newcommand{\ket}[1]{| #1 \rangle}

\newcommand{\braOket}[3]{\langle \: #1 \: | \: #2 \:| \: #3 \: \rangle}

\makeatletter
\newcommand*{\rom}[1]{\expandafter\@slowromancap\romannumeral #1@}
\makeatother
\begin{document}

\title{Attosecond transient absorption of  \\ a bound wave packet coupled to a smooth continuum}

\author{Jan~Marcus \surname{Dahlstr\"om}}
\email{marcus.dahlstrom@matfys.lth.se}
\affiliation{Department of Physics, Lund University, Box 118, SE-221 00 Lund, Sweden}
\affiliation{Department of Physics, Stockholm University, AlbaNova University Center, SE-106 91 Stockholm}

\author{Stefan \surname{Pabst}}
\affiliation{ITAMP, Harvard-Smithsonian Center for Astrophysics, 60 Garden Street, Cambridge, MA 02138, USA}
\affiliation{Stanford PULSE Institute, SLAC National Accelerator Laboratory, Menlo Park, California 94025, USA}

\author{Eva \surname{Lindroth}}
\affiliation{Department of Physics, Stockholm University, AlbaNova University Center, SE-106 91 Stockholm, Sweden}

%32.80.-t : Photoionization and excitation
%32.80.Ee : Rydberg states
%31.15.vj : Electron correlation calculations for atoms and ions: excited states
%42.65.Re : Ultrafast processes; optical pulse generation and pulse compression 
%32.80.Rm, 32.80.Qk, 42.65.Ky
%\date{}

\begin{abstract}
We investigate the possibility to use transient absorption of a coherent bound electron wave packet in hydrogen as an attosecond pulse characterization technique. In recent work we have shown that photoionization of such a coherent bound electron wave packet opens up for pulse characterization with unprecedented temporal accuracy --- independent of the atomic structure --- with maximal photoemission at all kinetic energies given a wave packet with zero relative phase [Pabst and Dahlstr\"om, {\it Phys. Rev. A}, {\bf 94}, 13411 (2016)]. Here, we perform numerical propagation of the time-dependent Schr\"odinger equation and analytical calculations based on perturbation theory to show that the energy-resolved maximal absorption of photons from the attosecond pulse does {\it not} uniquely occur at zero relative phase of the initial wave packet. Instead, maximal absorption occurs at different relative wave packet phases,  distributed as a non-monotonous function with a smooth $-\pi/2$ shift across the central photon energy (given a Fourier-limited Gaussian pulse). Similar results are found also in helium. Our finding is surprising because it implies that the energy-resolved photoelectrons are not mapped one-to-one with the energy-resolved absorbed photons of the attosecond pulse.       
\end{abstract}

\maketitle 

% Introduction:
\section{Introduction}
\label{sec:intro}
Ultra-short optical laser pulses provide the required time-resolution for femtochemistry: the study of molecular motion and transition states \cite{Zewail2000}, but optical pulses can not be made short enough to probe valence electron motion in atoms and molecules, which typically occur on the attosecond, or few femtosecond, time scale. Today it is possible to reach the electron time scale using attosecond pulses of coherent extreme ultraviolet (XUV) or x-ray radiation, which has opened up for attophysics: the study electron motion in atoms and molecules \cite{Krausz2009}. 
As attophysics is the natural continuation of femtochemisty, it is not surprising that many experimental ideas have already been transfered from the femtosecond to the attosecond domain. One example of such a technique is attosecond transient absorption spectroscopy (ATAS), which has been used to study electron coherences and to reconstruct the motion of valence electrons in atoms \cite{GouliemakisNature2010,Wirth2011,Sabbar2017}. With increasing photon energies, time-resolved x-ray absorption spectroscopy (TR-XAS) makes it possible to follow element-specific dissociation of molecules in real time  \cite{Pertot2017}. In the above mentioned experiments an intense laser field was used to prepare a wave packet in the target system, while the coherent XUV/x-ray light was used as a probe of the dynamics. The reversed scenario is also possible, where, instead, the attosecond pulse is used to trigger electron dynamics that are sequentially probed with the strong laser field that couples populated autoionizing states to dark states. This gives rise to numerous effects including control of the Fano line shape, light-induced states and  Autler-Townes splitting \cite{Wang2010,OttScience2013,OttNature2014} and to the  observation of counter-rotating wave effects \cite{PhysRevA.95.053420}. 
An overview of experimental work with ATAS is given in Ref.~\cite{BECK2015119}.

The ATAS method has been studied extensively in theory with the main focus 
on the description of dynamics of laser-coupled autoionizing states \cite{Chu2013} and 
on the pulse overlap (non-sequential) region \cite{Pabst2012}. The theory of strong-field ATAS is comprehensively reviewed in Ref.~\cite{WuJPB2016}, but the simple case of ATAS sequentially coupled to a smooth (resonance free) continuum by an attosecond pulse has not yet been considered. The question arises if autoionizing states are required to control absorption of the attosecond pulse or if it is possible to exert similar control when ATAS is coupled to a smooth continuum? 

In the light of our recent proposal, where we showed that sequential photoionization of a bound wave packet  can be used as a characterization method for attosecond pulses with unprecedented temporal accuracy \cite{PabstPRA2016,PabstJPB2017}, we ask the question: can ATAS serve as an all-optical pulse characterization method? The article is organized as follows. In Sec.~\ref{sec:theory} we outline the theory of our method including (A) information about our numerical propagation scheme and (B) analytical calculations based on lowest-order perturbation theory, in Sec.~\ref{sec:results} we present our main results and in Sec.~\ref{sec:conclusions} we present our conclusions.

\section{Theory}
\label{sec:theory}
We write the time-dependent Schr\"odinger equation (TDSE) for a single electron as
\begin{align}
i\ket{\dot \psi(t)}=
\hat H(t) \ket{\psi(t)}=
\left[\hat H_0 + \hat V_I(t)\right]\ket{\psi(t)}, 
\label{TDSE}
\end{align}
where the total Hamiltonian, $\hat H$, consists of a time-independent atomic operator, $\hat H_0$, and a time-dependent field-interaction operator, 
\begin{align}
\hat V_I(t)=A(t) \hat p_z+\frac{1}{2}A^2(t). 
\label{VI}
\end{align}
The field-interaction in this form assumes a linearly polarized vector potential, $A(t)$, along the $z$-axis within the velocity gauge dipole approximation with the z-component of the momentum operator  denoted as $\hat p_z$. The eigenstates of the atomic Hamiltonian are denoted by $\ket{j}$ with eigenvalues, $\epsilon_j$. We use short-hand notation $\ket{j}$ to label all states with atomic quantum numbers for bound states $(n_j,\ell_j,m_j)$ and energy-normalized continuum states $(\epsilon_j,\ell_j,m_j)$. Atomic units are used unless otherwise stated, $e=\hbar=m_e=4\pi\epsilon_0=1$. 

\subsubsection{Gauge consideration}
\label{sec:gauge}
In the following we will refer to the vector potential in Eq.~(\ref{VI}) as the {\it test pulse}. It is common practice to remove the interaction term proportional to $A^2(t)$. This omission can be fully justified \cite{scrinzi:2010}, also in cases then the test pulse is strong, by performing a spatially uniform, time-dependent gauge transformation,  $\ket{\psi}=\exp(i\Lambda)\ket{\tilde\psi}$, where
\begin{align}
\Lambda(t) = -\int^t dt' \frac{1}{2}A^2(t'). 
\label{Lambda}
\end{align}
While this unitary transformation leaves the expectation value of the canonical momentum operator unchanged, 
\begin{align}
p_z(t)=\braOket{\psi}{\hat p_z}{\psi}=\braOket{\tilde \psi}{\hat p_z}{\tilde \psi}, 
\end{align}
it does involve a subtle re-definition of the absolute energy scale at each moment in time that we need to be aware of in the following work. Finally we add that the expectation value of the position operator in length gauge $(L)$ is identical to that computed using the wave functions from velocity gauge $(V)$,
\begin{align}
z(t)=
\braOket{\psi^{(L)}}{\hat z}{\psi^{(L)}}=
\braOket{\psi^{(V)}}{\hat z}{\psi^{(V)}},
\label{zLzV}
\end{align}
which can be shown with the associated well-known gauge transformation, 
\begin{align}
\Lambda'(z,t)=-zA(t).
\end{align}

\subsubsection{Initial bound wave packet and model test pulse}
\label{sec:initial}
We will consider that the initial condition for the TDSE in Eq.~(\ref{TDSE}) is a field-free wave packet consisting of $N$ bound states labeled by $j$,
\begin{align}
\ket{\psi^{(0)}(t)}=\sum_j^N c_j(t) \exp[-i\epsilon_j t]\ket{j},
\end{align}
where $c_{j}(t)$ are complex amplitudes that are constant under free-field conditions (before the test pulse arrives on target). The scheme for preparing the wave packet is not important, but it is essential that it is sequential. This means that the preparation of the wave packet is finished before the test pulse arrives at the atom. It goes without saying that the wave packet is assumed to be fully coherent with the test pulse. 

As mentioned in the introduction, the role of excitation to (quasi) bound states has been studied recently by isolated attosecond pulses.  In this paper, we study the case where the test pulse ionizes the atom, i.e. the central frequency is larger than the ionization potential: $\omega_0 > I_p$, with excitation to a smooth continuum of states. As a concrete example we will consider the following two-state wave packet and test pulse: 
\begin{align}
\mathrm{Wave packet:}&
\left\{
\begin{array}{lll}
c_1&=c_2&=1/\sqrt{2} \\
\epsilon_{j_1} &= 0.375 \, %\mathrm{au}
&= 10.20 \,\mathrm{eV} 
\\
\epsilon_{j_2} &= 0.444 \, %\mathrm{au}
&= 12.09 \,\mathrm{eV}  \\ 
I_p &= 0.5 \, %\mathrm{au}  
&= 13.61 \,\mathrm{eV} \\
\end{array}
\right. \nonumber \\
\mathrm{Test \,pulse:}&
\left\{
\begin{array}{lll}
\omega_0    &= 1.5\,%\mathrm{au} 
&= 40.81 \,\mathrm{eV} 
\\
\tau_e &= 5.0\,%\mathrm{au} 
&= 120.94 \,\mathrm{as}, 
\end{array}
\right.
\label{paperparam}
\end{align}
where the atomic energies correspond to a $2s/3s$-wave packet in the hydrogen atom. The vector potential is modeled by an oscillating  Gaussian pulse, 
\begin{align}
A(t)=A_0\cos[\omega_0( t-t_0)+\phi]
\exp\left[-a (t-t_0)^2\right],
\label{vectorpotential}
\end{align}
where $A_0$ is the envelope amplitude, $\tau_0$ is the arrival time of envelope, $\phi$ is the phase shift of the oscillations and $a=2\ln(2)/\tau_e^2$ is inversely proportional to the squared envelope duration, $\tau_e$, expressed in full-width at half-maximum (FWHM).

\subsubsection{Energy absorbed by atom}
\label{sec:energyabsorbed}
Due to total energy conservation it is possible to infer the absorbed energy from the test pulse by instead calculating the energy gained by the excited atom, $\Delta {\cal  E }$. This latter quantity can be formally expressed as \cite{WuJPB2016}
\begin{align}
\Delta {\cal E} = \int_{-\infty}^{+\infty}dt\frac{d\Delta {\cal E}}{dt},
\label{DE}
\end{align}
where $\Delta {\cal E}(t)$ is the time-dependent expectation value of the atomic energy in the field. The instantaneous power delivered to the atom in velocity gauge becomes 
\begin{align}
\frac{d\Delta { \cal E}}{dt}=&\frac{d}{dt}\braOket{\psi(t)}{\hat H(t)}{\psi(t)} \nonumber \\ =&
\left[ p_z(t)+A(t) \right] \dot A(t) = -\Pi_z(t)E(t),
\label{dDEdt}
\end{align}
%where $p_z(t)=\braOket{\psi(t)}{\hat p_z}{\psi(t)}$ is the time-dependent expectations value of $\hat p_z$ and $\dot A(t)=\partial A / \partial t$ is the time derivative of the test pulse. 
which is gauge invariant because it involves the kinetic momentum, $\Pi_z=p_z+A$, and the electric field, $E=-\dot A$. Eq.~(\ref{dDEdt}) is  analogous to the expression found in classical electrodynamics when computing the power transfered to a classical system. If, instead, we were to use the gauge transformed Hamiltonian (without the $\frac{1}{2}A^2(t)$ term) then the $A\dot{A}$ term in Eq.~(\ref{dDEdt}) would be missing and it would be the canonical momentum that is multiplied with the electric field. Clearly, this power is then not gauge invariant because it must be related to the time-dependent absolute energy associated with the unitary transformation in Eq.~(\ref{Lambda}), but the total energy gain is gauge invariant because   
\begin{align}
\int_{-\infty}^{+\infty}dt A(t)\dot{A}(t)=\left.\left[\frac{1}{2}A^2(t)\right]\right|_{-\infty}^{+\infty}=0,
\label{intAdotA}
\end{align}
provided that the test pulse is finite in time. 
The $A\dot{A}$ term is universal because it does not involve any reference to the specific atomic system under investigation. Instead, it can be interpreted as the power delivered to a free electron with no initial or final momentum, but with a field-induced momentum, $A(t)$, during the action of the test pulse. In this interpretation it is clear that the integrated power (total energy gain) mediated by the $A\dot{A}$ term must vanish as a free electron should be unable to gain energy from any given test pulse.    

Following Ref.~\cite{WuJPB2016}, but instead using velocity gauge, the total energy gain in Eq.~(\ref{DE}) can be rewritten in the energy domain as 
\begin{align}
\Delta {\cal E} = &
%\int_{-\infty}^{+\infty} d\omega i\omega \tilde p_z(\omega) \tilde A^*(\omega)  \\  = &
-2\int_{0}^{+\infty} d\omega \,
\omega \mathrm{Im}\left\{
 [ \tilde p_z(\omega) + \tilde A(\omega) ] \tilde A^*(\omega) 
\right\}, 
\label{DEfreq}
\end{align}
where $\tilde p_z(\omega)=\tilde p_z^*(-\omega)$ and $\tilde A(\omega)=\tilde A^*(-\omega)$ are the Fourier transforms of the real $p_z(t)$ and $A(t)$ functions, respectively. This implies that the energy-resolved  absorbed energy from the test pulse (or gained by the atom) is
\begin{align}
\frac{
d\Delta {\cal E}
}{d\omega}
= -2\omega\mathrm{Im}
\left\{\tilde p_z(\omega) \tilde A^*(\omega)\right\},
\label{dDEdw}
\end{align}
where we have used the fact that the term $\tilde A \tilde A^*=|\tilde A|^2$ in Eq.~(\ref{DEfreq})
is real. We find that the $A \dot A$ term in Eq.~(\ref{dDEdt}), does not contribute to absorption of the test pulse at {\it any} energy. In other words, the energy density in Eq.~(\ref{dDEdw}) is invariant with respect to the unitary transformation given by Eq.~(\ref{Lambda}), in contrast to the power given in Eq.~(\ref{dDEdt}).  

\subsection{Numerical propagation}
\label{sec:numprop}
In this section we discuss our numerical approach to study absorption of a test pulse by a coherent wave packet in hydrogen and noble gas atoms, such as helium. The effective radial Schr\"odringer equation is
\begin{align}
\left[
- \frac{1}{2}\frac{\partial^2}{\partial r^2} 
+ \frac{1}{2}\frac{\ell(\ell+1)}{r^2}
- \frac{Z}{r} 
+ u_\mathrm{HF}
\right]\psi_j(r)=\epsilon_j\psi_j(r),
\end{align}
where $Z$ is the nuclear charge and $u_\mathrm{HF}$ is the  Hartree-Fock potential (for hydrogen we have $Z=1$ and $u_\mathrm{HF}=0$).
We use a primitive basis set of B-splines \cite{R1980} to find the diagonalized hydrogen states, or restricted Hartree-Fock (HF) orbitals \cite{mbpt}, using codes developed earlier in Stockholm for time-independent calculations, c.f Ref.~\cite{amp:82:sms,DahlstromPRA2012,DahlstromJPB2014}. The eigenstates are labeled by $j$ for a given electron state and expressed in non-orthogonal B-splines labeled by $n$,  
\begin{align}
\psi_j(r)=\sum_n^{N_b} b_n^{(j)} B_n(r),
\label{phiinB}
\end{align}
with complex amplitudes $b_n^{(j)}$. We use B-splines of order $k=5$ with a grid of $N_p=220$ knot points, thus, generating a total number of $N_b=N_p-k-2=213$ B-splines that vanish at the radial boundaries of the grid. The knot points are placed in the complex plane in accordance with exterior complex scaling (ECS) \cite{SimonPLA1979}, 
\begin{align}
r \rightarrow
\left\{
\begin{array}{lr}
r, & 0 < r  < R_C\\
R_C + \left(r-R_C\right) e^{i \theta}, & r > R_C
\end{array}
\right.
\label{ecs}
\end{align}
where the radial coordinate is transformed beyond a radius $R_C=75$ with complex angle $\theta=15^o$. 
A wave packet expressed as a superposition of eigenstates of the scaled Hamiltonian is physical in the inner region, $r<R_C$, while it is exponentially damped (unphysical) in the outer region, $r>R_C$. We place 17 knot points in this outer region to account for the damping from $75$ to $90$ along the real axis. ECS is a useful tool for numerical propagation as it avoids numerical reflections at the end of the numerical box, c.f. Ref.~\cite{Scrinzi2010} where the complex scaling is done all the way to infinity. 

The TDSE is propagated on the HF basis similar to the time-dependent configuration-interaction singles (TDCIS) implementation presented in Refs.~\cite{Greenman2010,Pabst2012}, but here we are restricted to propagation in the velocity gauge $(V)$ due to ECS \cite{McCurdy1991}. None the less, it is possible to compute the time-dependent dipole also in length gauge $(L)$ using the ``physical'' $(V)$-wave packet in the inner region, as discussed in connection with Eq.~(\ref{zLzV}). In practice this is done easily by truncating the summation of B-splines [Eq.~(\ref{phiinB})] when constructing the operators so that only unscaled (real) B-splines are used to compute the radial part of the dipole matrix elements. We have verified that changing the position of the truncation does not alter the results presented in this paper. 

\begin{figure}[!htb]
\includegraphics[width=0.5\textwidth]{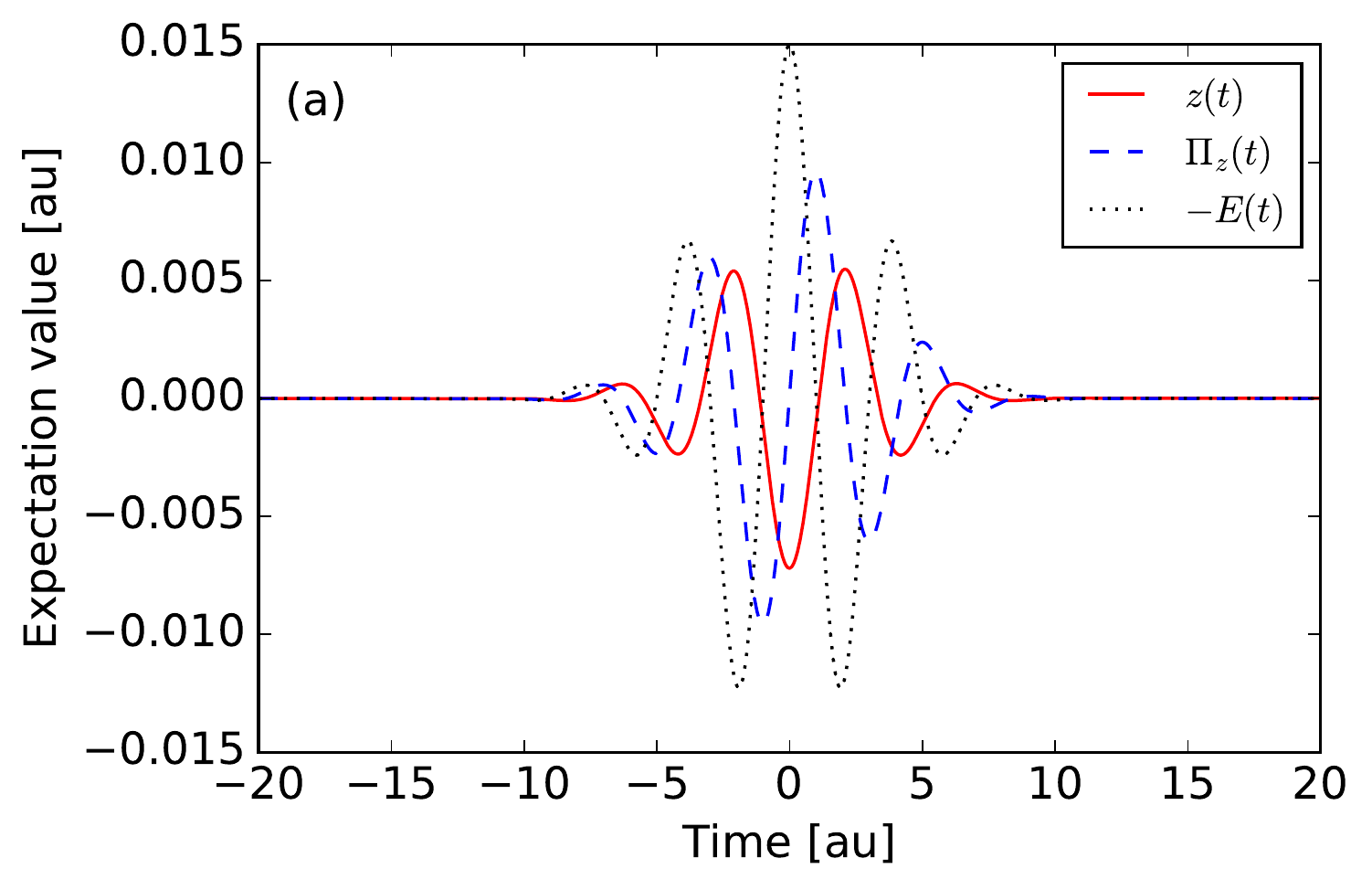}
\includegraphics[width=0.5\textwidth]{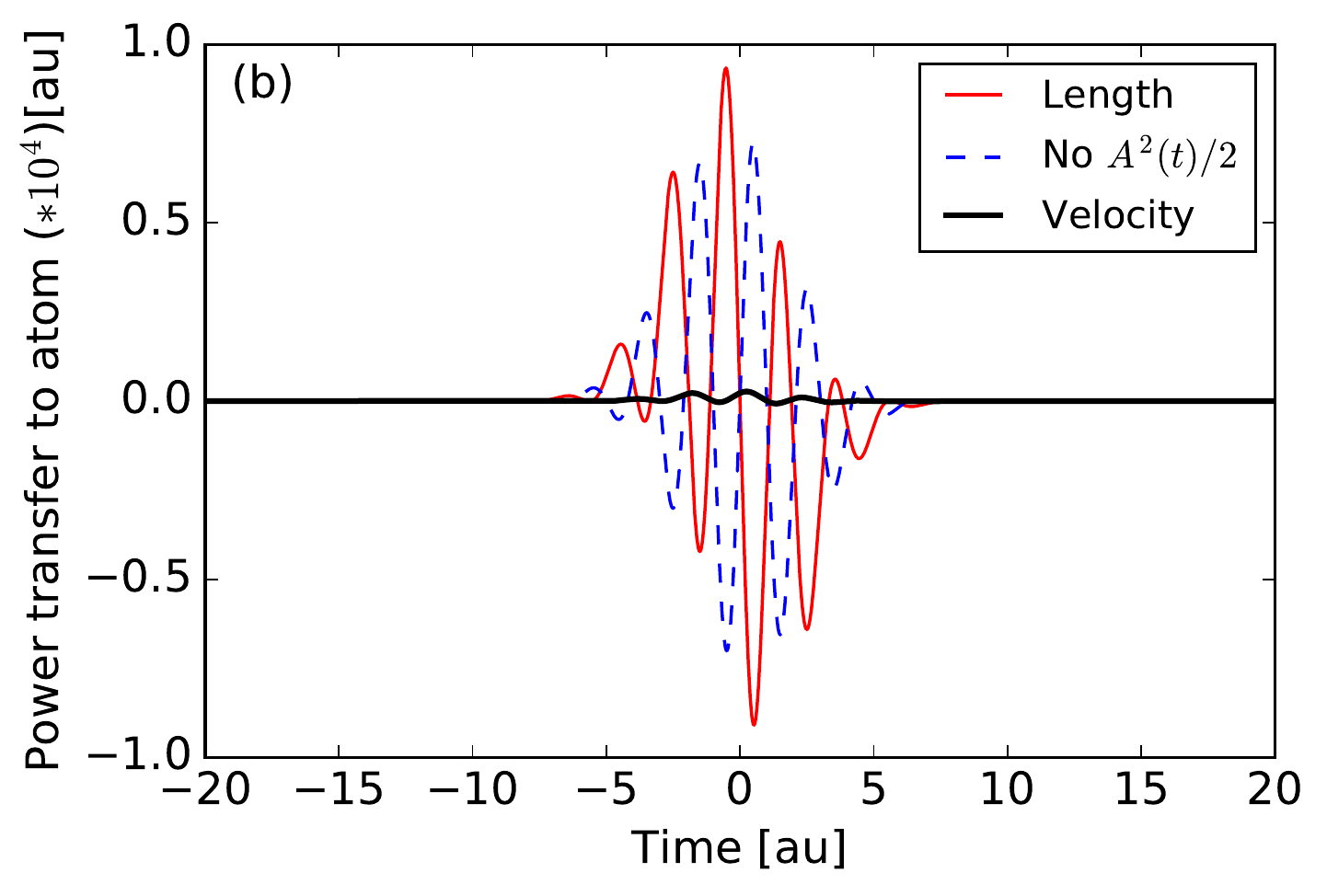}
\includegraphics[width=0.5\textwidth]{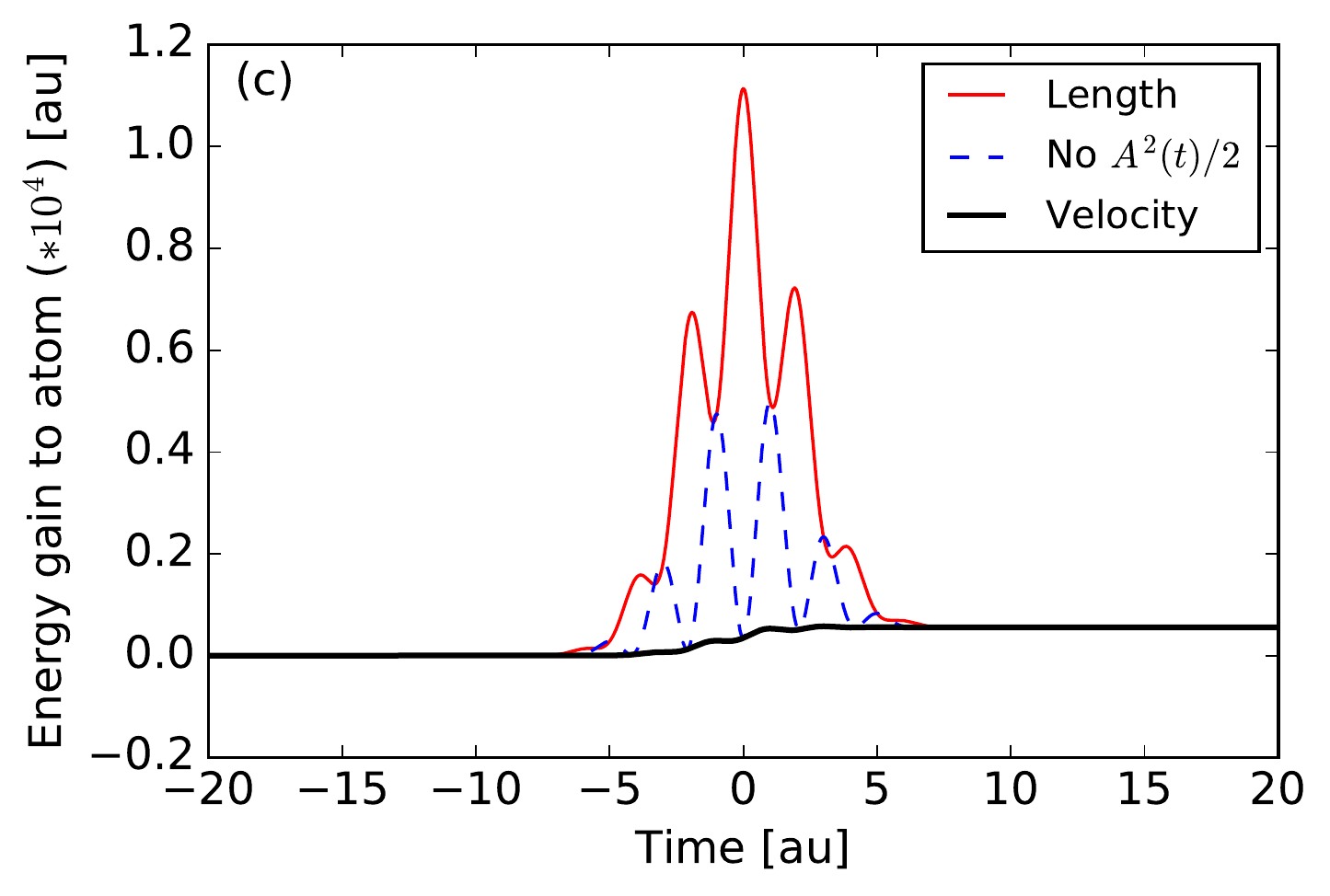}
\caption{
(a) Time-dependent expectation values of position and kinetic momentum induced by test pulse starting with a $2s/3s$ hydrogen wave packet. The force (minus electric field) of the test pulse is plotted for direct comparison. 
(b) Power delivered to the atom as a function of time.
(c) Energy delivered to the atom as a function of time. See main text for details about the three different gauges.  
\label{fig:numprop}}
\end{figure}

\subsubsection{Hydrogen: 2s/3s wave packet}
\label{sec:H}
We now consider the initial $2s/3s$ wave packet in hydrogen and pulse parameters shown in Eqs.~(\ref{paperparam})-(\ref{vectorpotential}) with a weak field strength $A_0=0.01$. In Fig.~\ref{fig:numprop}~(a) we show the corresponding time-dependent expectation values of position and kinetic momentum of the electron. The electron oscillates during the test pulse, see $z(t)$, with a phase lag of $\sim\pi$ relative the driving force of the test pulse, $-E(t)$. In analogy with a damped simple harmonic oscillator this $\pi$-shift implies a strongly over-driven system, i.e. the electron has a hard time to keep up with the fast oscillations of the test pulse. The kinetic momentum, $\Pi_z(t)$, is shifted by $\sim \pi/2$ in between force and position, as expected for an oscillating system. The expectation values oscillate only during the overlap region with the test pulse. The expectation values of an unprepared hydrogen atom (in the $1s$ state) looks qualitatively the same (not shown).    

In Fig.~\ref{fig:numprop}~(b) we show the  instantaneous power delivered to the atom according to Eq.~(\ref{dDEdt}) in velocity gauge, both with and without the $\frac{1}{2}A^2$ term, as well as the length-gauge expression used by Wu et al. in Ref.~\cite{WuJPB2016}, 
\begin{align}
\Delta \dot{{\cal E}}^{(L)}(t)=z(t)\dot{E}(t). 
\label{DdotEL}
\end{align}
The power delivered to the atom during the action of the test pulse is gauge dependent. In length gauge and the transformed velocity gauge (without $\frac{1}{2}A^2(t)$), the power oscillates between positive and negative values, which implies that power is delivered and extracted at a fast rate during the test pulse. At the center the test pulse the power oscillations are $\pi$-shifted with respect to length and transformed velocity gauge, which shows that the interpretation of these quantities as ``power'' is debatable. In velocity gauge the power looks dramatically different: It is purely positive and much smaller in magnitude. 
It is known from text books on quantum mechanics, c.f. Ref.~\cite{sakurai:modern}, that the absolute energy scale is shifted by a spatially uniform time-dependent gauge transform, such as $\frac{1}{2}A^2(t)$. This explains the difference between the transformed and original velocity gauge power because the time-dependent redefinition of the absolute energy scale can be taken into account by adding $-A(t)E(t)$ to the transformed velocity gauge result giving back the velocity result. From these considerations we consider the velocity gauge as the correct ``physical'' convention for the instantaneous power.   

In Fig.~\ref{fig:numprop}~(c) we show the energy gain (integrated power) delivered to the atom as a function of time. Again, during the action of the test pulse, the energy gain is gauge dependent. The length gauge and the transformed velocity gauge show large oscillations of the energy gain, while the physical energy delivered (velocity gauge) shows that the energy is gradually increased in small steps toward the final value. After the test pulse is over, however, the total energy gain is identical in all three gauges --- which shows that either gauge can be used to compute the total energy gained by the test pulse.  

\begin{figure}[!htb]
\includegraphics[width=0.5\textwidth]{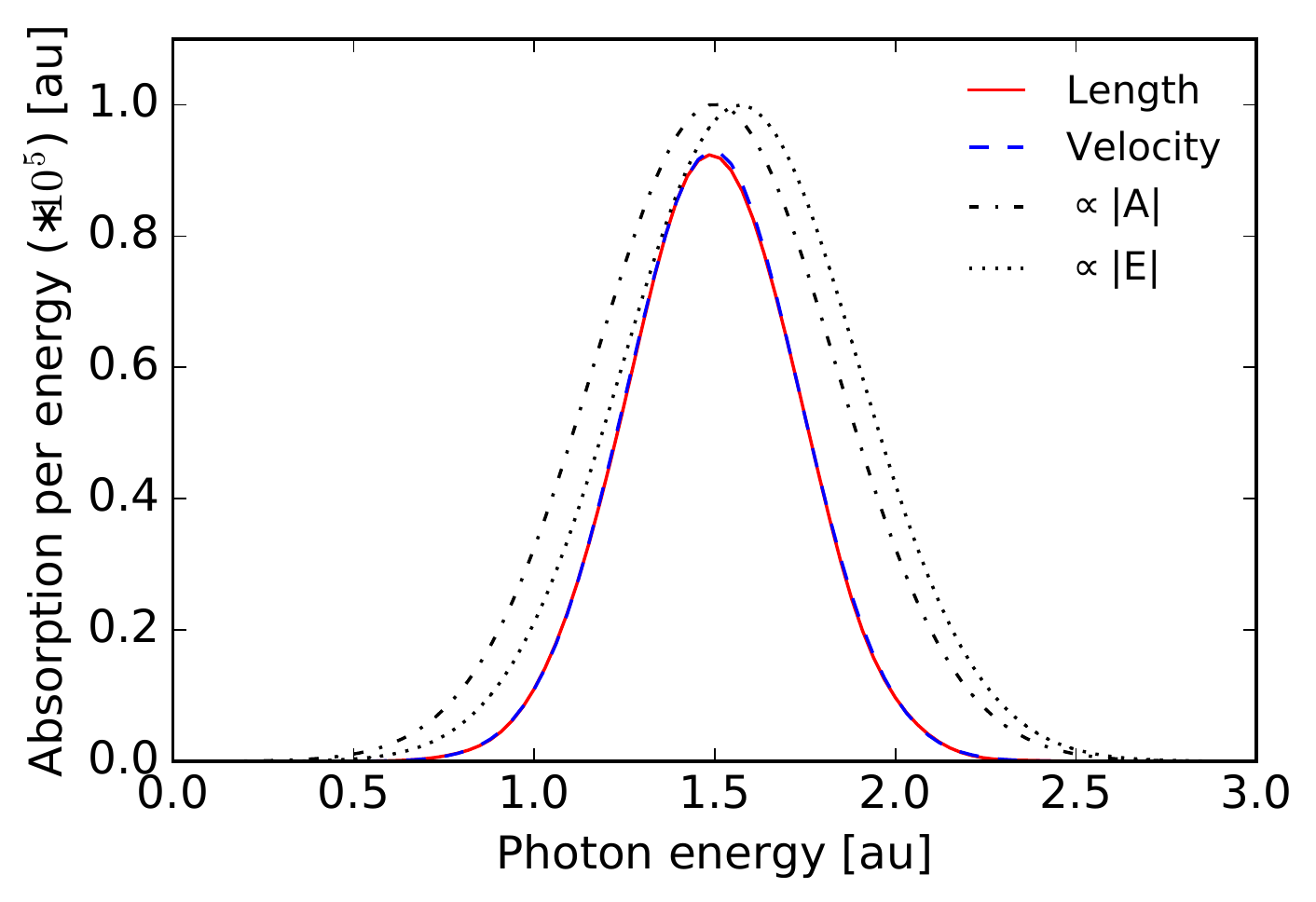}
\caption{Energy absorbed by the hydrogen atom, initially in a 2s/3s wave packet, resolved over photon energies. The normalized electric field and vector potential are shown for comparison in the energy domain. The data has been interpolated between discrete values by a cubic spline function. 
\label{fig:dDEdw}}
\end{figure}

In Fig.~\ref{fig:dDEdw} we show the energy absorbed by the 2s/3s wave packet in the hydrogen atom resolved in photon energy evaluated by Eq.~(\ref{dDEdw}). 
The normalized Fourier transform of the electric field and vector potential are shown for comparison. The peak of absorption is located close to the central photon frequency, $\omega_0=1.5$, in good agreement with the peak of the vector potential, while the electric field peak is located at a slightly higher energy. Finally, we note that the energy resolved absorption is gauge invariant for the hydrogen wave packet. 

In Table~\ref{tab:abs} we show the total absorption of energy from the first three $s$-wave states in hydrogen individually. The absorption is decreasing with increasing principal quantum number and the ratio of 2s/3s is 3.45. 
\begin{table}
\begin{tabular}{|l|l|}
\hline Initial state: & Absorption [au]: \\ \hline 
$1s$ & 6.87e-05 \\
$2s$ & 7.38e-06 \\
$3s$ & 2.14e-06 \\ \hline
\end{tabular}
\caption{
\label{tab:abs}
Absorbed energy by the hydrogen atom prepared in different initial states before the arrival of the test pulse. 
}
\end{table}

The question then arises as to how a delay of the coherent 2s/3s wave packet can change the absorption of the test pulse. In order to study this effect we assume that 
\begin{align}
c_2= c_1\exp[i\varphi]
\label{varphase}
\end{align}
where $\varphi$ is the relative phase of the initial wave packet. 
\begin{figure}[!htb]
\includegraphics[width=0.5\textwidth]{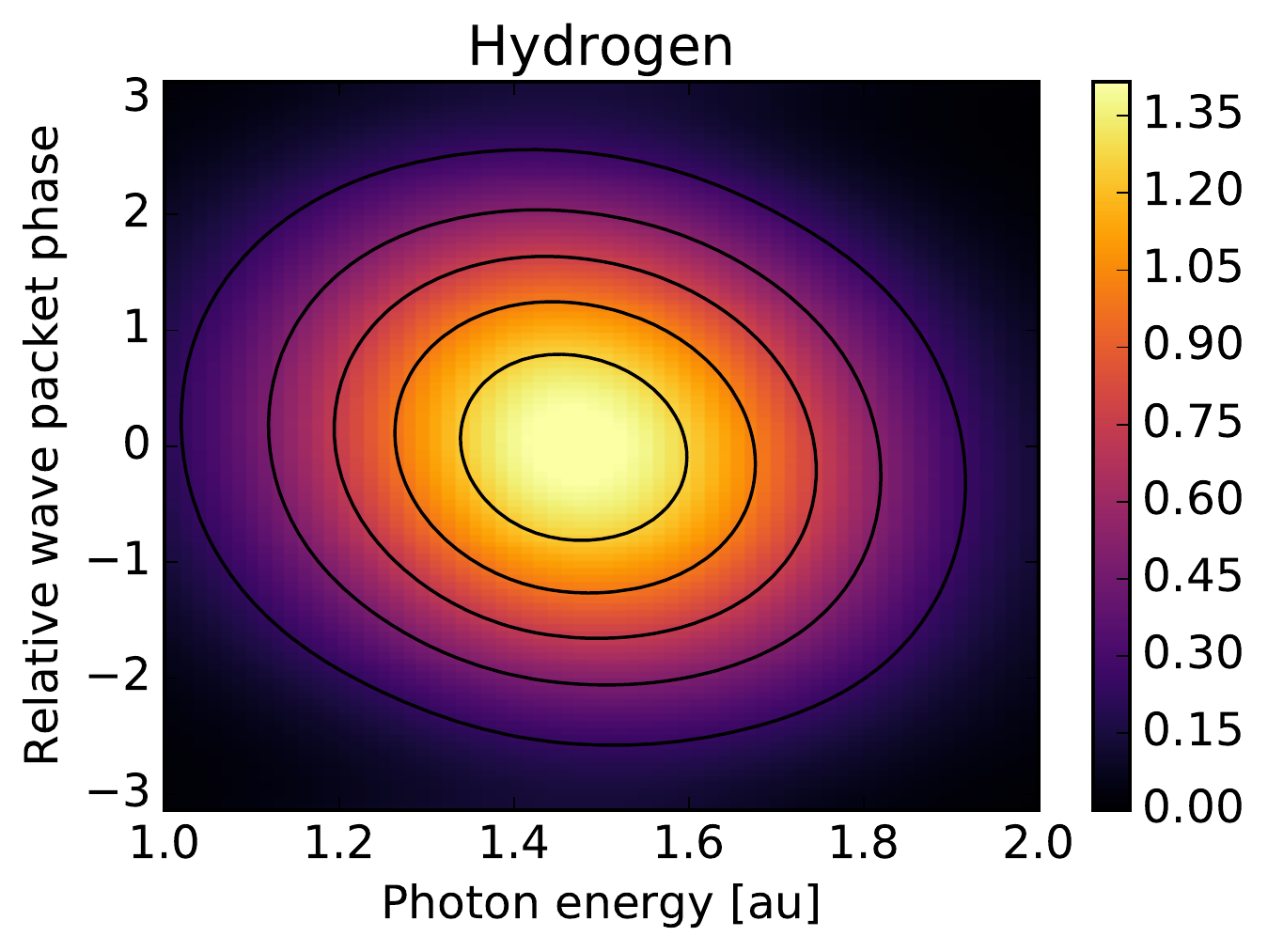}
\caption{The absorbed energy ($*10^5$) from the test pulse by the $2s/3s$ hydrogen wave packet as a function of photon energy and relative wave packet phase, $\varphi$. The data has been interpolated between discrete values by a cubic spline function.  
\label{fig:Hmap}}
\end{figure}
In Fig.~\ref{fig:Hmap} we show the energy-resolved absorption as a function of $\varphi$  in false color. The absorption of energy from the test pulse is most efficient if the wave packet states are in phase with each other, $\varphi\approx 0$. Conversely, if the wave packet states have opposite phases, $\varphi\approx\pi$, then the absorption is reduced. The atom can not be made completely  transparent for the test pulse, but the energy gain can be dramatically reduced by almost an order of magnitude. Further, there is a tilt in the absorption map, where the low (high) photon energies require a slightly larger (smaller) relative phase to maximize the absorption. This observation is surprising given our recent work on pulse characterization using {\it photoemission} from coherent bound wave packets, referred to as pulse analysis by delayed absorption (PANDA) \cite{PabstPRA2016,PabstJPB2017}. When using the PANDA method, the total photoemission (photoelectrons collected over all emission angles) from an identical bound $2s/3s$ wave packet in hydrogen, would have maximal ionization signal at all kinetic energies exactly at the zero relative phase of the wave packet ($\varphi=0$). Here, we expected a similar effect to occur in ATAS, but Fig.~\ref{fig:Hmap} shows that this is clearly not the case. 
In order to understand the process of absorption from a bound wave packet in more detail, we must turn to perturbation theory. 

\subsection{Perturbation theory}
\label{sec:pt}
The energy absorbed by the atom can be worked out analytically provided that the test pulse is weak and has a simple shape. If we assume that the complex amplitudes of the bound wave packet are constant, 
\begin{align}
c_j(t) \approx c_j^{(0)}, 
\label{cj}
\end{align}
equal to their initial value before the arrival of the test pulse, then the first-order wave packet generated by the test pulse is
\begin{align}
\ket{\psi^{(1)}(t)}=\sumint_f c_f(t) \exp[-i\epsilon_f t] \ket{f},
\end{align}
with complex amplitudes 
\begin{align}
c_f(t) = \frac{1}{i}\sum_j\braOket{f}{\hat p_z}{j}c_j
\int_{-\infty}^{t}dt' A(t') \exp[i\omega_{fj}t'], 
\label{cf}
\end{align} 
where $\omega_{fj}=\epsilon_f-\epsilon_j$. 
Given the test pulse in Eq.~(\ref{vectorpotential}), the complex amplitudes for the excited wave packet can be written as
\begin{align}
c_f(t)=\frac{A_0}{2i}\sum_j\braOket{f}{\hat p_z}{j}c_j\left[ F_+(t)+F_-(t) \right],
\end{align}
where the time-dependent factors can be evaluated in terms of error functions,  
\begin{align}
F_\pm(t)=&
e^{\pm i \phi}
\int_{-\infty}^{t}d\tau
\exp\left[
i(\omega_{fj}\pm\omega_0)\tau-a\tau^2]
\right]
\nonumber \\&=-
\frac{e^{\pm i \phi}}{2}\sqrt{\frac{\pi}{a}}
\exp\left[-\frac{(\omega_{fj}\pm\omega_0)^2}{4a}\right]
\nonumber \\ & \times  \left[\mathrm{erf}
\left(
\frac{2at-i(\omega_{fj}\pm\omega_0)}{2\sqrt{a}}
\right)
+1
\right].
\label{Fpm}
\end{align}
%In Eq.~(\ref{Fpm}) $\mathrm{erf}(z)$ is 
The error function tends to $\pm$one as time goes to $\pm$infinity, which implies that the last factor in Eq.~(\ref{Fpm}) is zero before the test pulse and  two after the test pulse. In general the $F_-(t)$ contribution is larger than the $F_+(t)$ contribution because of the resonance condition with the continuum:  
$\omega_{fj}=\epsilon_f-\epsilon_j\approx\omega_0$. 
The omission of the $F_+(t)$ term is called the rotating wave approximation (RWA).   

The time-dependent expectation value of $\hat p_z$ is approximated using the zeroth- and first-order wave packet as
\begin{align}
p_z(t)\approx & 
\bra{\psi^{(0)}(t)} \hat p_z \ket{\psi^{(1)}(t)} + c.c.
\nonumber \\ =&
\frac{A_0}{2i}\sum_{j',j}\sumint_f
c_{j'}^* c_j
\braOket{j'}{\hat p_z}{f}
\braOket{f}{\hat p_z}{j}
\nonumber \\ & \times 
\exp[-i\omega_{fj'}t]
\left\{
F_+(t)+F_-(t)
\right\}+c.c.,
\label{pzt}
\end{align}
where we can clearly see that the electron makes a transition from any initial wave packet state $j=\{j_1,j_2\}$ to any (dipole-allowed) excited state $f$ and then back to any wave packet state $j'=\{j_1,j_2\}$ with amplitudes proportional to the Fourier transform of the time-factors at the recombination energy, 
$\omega_{fj'}=\epsilon_f-\epsilon_{j'}$.   
In the Fourier domain the expression is indeed  simplified to 
\begin{align}
\tilde p_z^\pm(\omega) = &
-\frac{A_0}{2\sqrt{2a}}
\sum_{j,j'}\sumint_f
c^*_{j'}c_j 
\braOket{j'}{\hat p_z}{f}
\braOket{f }{\hat p_z}{j} \nonumber \\
\times &
\left\{
\frac{\mathrm{p.v.}}{\omega-\omega_{fj'}}
-i\pi \delta(\omega-\omega_{fj'})
\right\}
\nonumber \\ 
\times &
\exp\left[
\pm i \phi
-\frac{(\omega\pm\omega_0-\omega_{jj'})^2}{4a}
\right], 
\label{pzw}
\end{align}
where we have obtained a non-resonant principal value  term and a resonant delta function term for each separate time factor $F_\pm(t)$ in Eq.~(\ref{pzt}). We see that the $p_z^\pm(\omega)$ expressions for plus and minus are equivalent besides the trivial phase factor of the field and the Gaussian field spectral envelope [last factor in Eq.~(\ref{pzw})]. The p.v.-term in Eq.~(\ref{pzw}) arises from the time-dependent population of the excited wave packet in Eq.~(\ref{Fpm}), while the delta term arises from the constant term. The p.v. term was derived by partial integration (boundary terms are neglected) using the fact that the primitive function of the error function is a Gaussian function in time that then transforms to a Gaussian function in energy. 

\subsubsection{Toy model: 2s/3s wave packet}
\label{sec:toymodel}
The evaluation of Eq.~(\ref{pzw}) requires knowledge of the transition matrix elements of the system. Thus, we introduce a simple toy model with transition matrix elements:
\begin{align}
\braOket{f}{\hat p_z}{j}=
\left\{
\begin{array}{ll}
\frac{1}{(1+k)^2},& \epsilon_f>I_p \\
0,& \epsilon_f<I_p,
\end{array}
\right.
\label{toymodel}
\end{align}
where $k=\sqrt{2(\epsilon_f-I_p)}$ is the wave number of the photoelectron. The toy model has two important features: (i) excited bound states are neglected as the transition matrix element is zero for photon energies below the ionization threshold, $I_p$; and (ii) the transition strength decreases at high photon energies with a characteristic rate of the hydrogen atom. If the termination of the integrand is not enforced, the p.v. integral in Eq.~(\ref{pzw}) may diverge. 
\begin{figure}[!htb]
\includegraphics[width=0.5\textwidth]{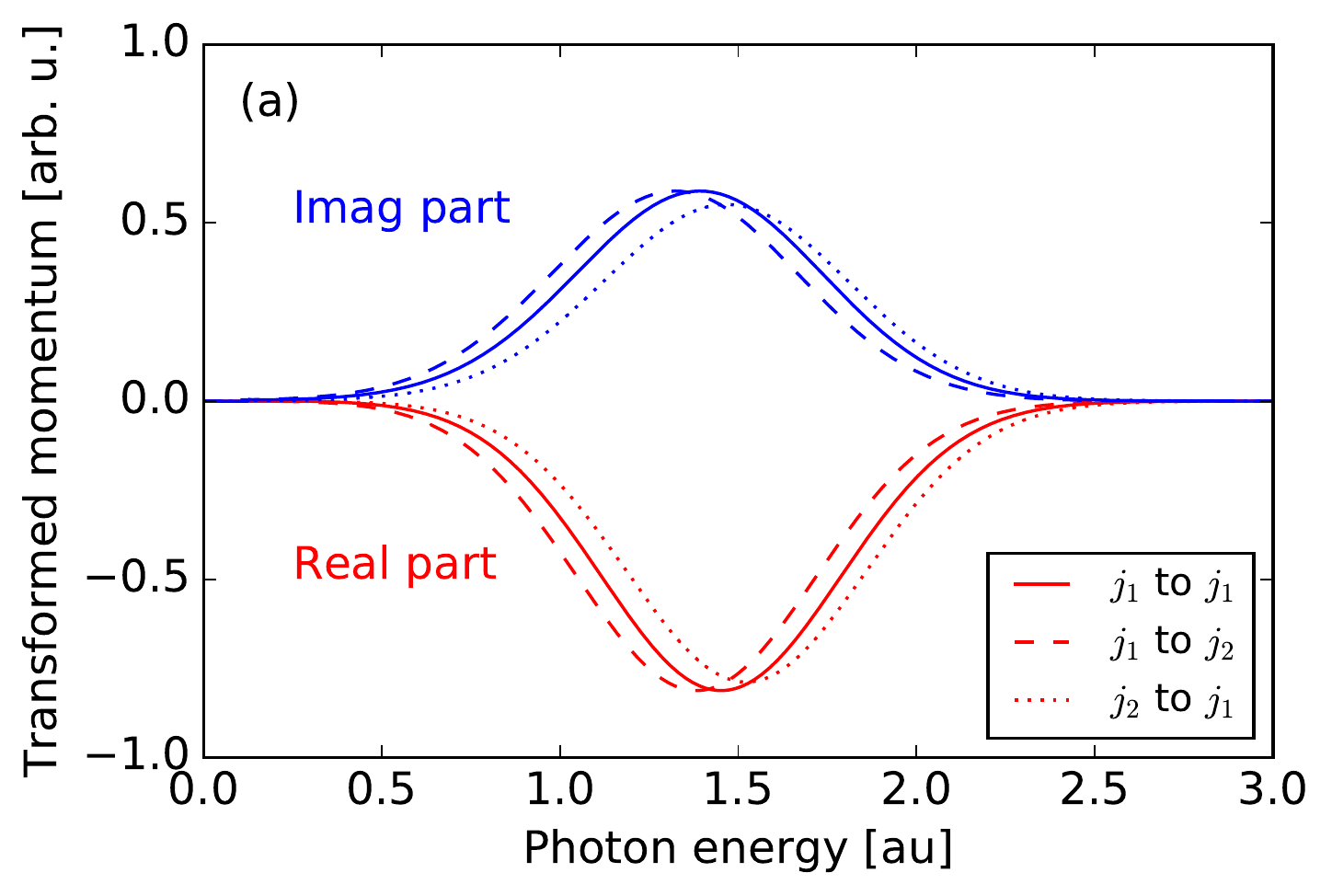}
\includegraphics[width=0.5\textwidth]{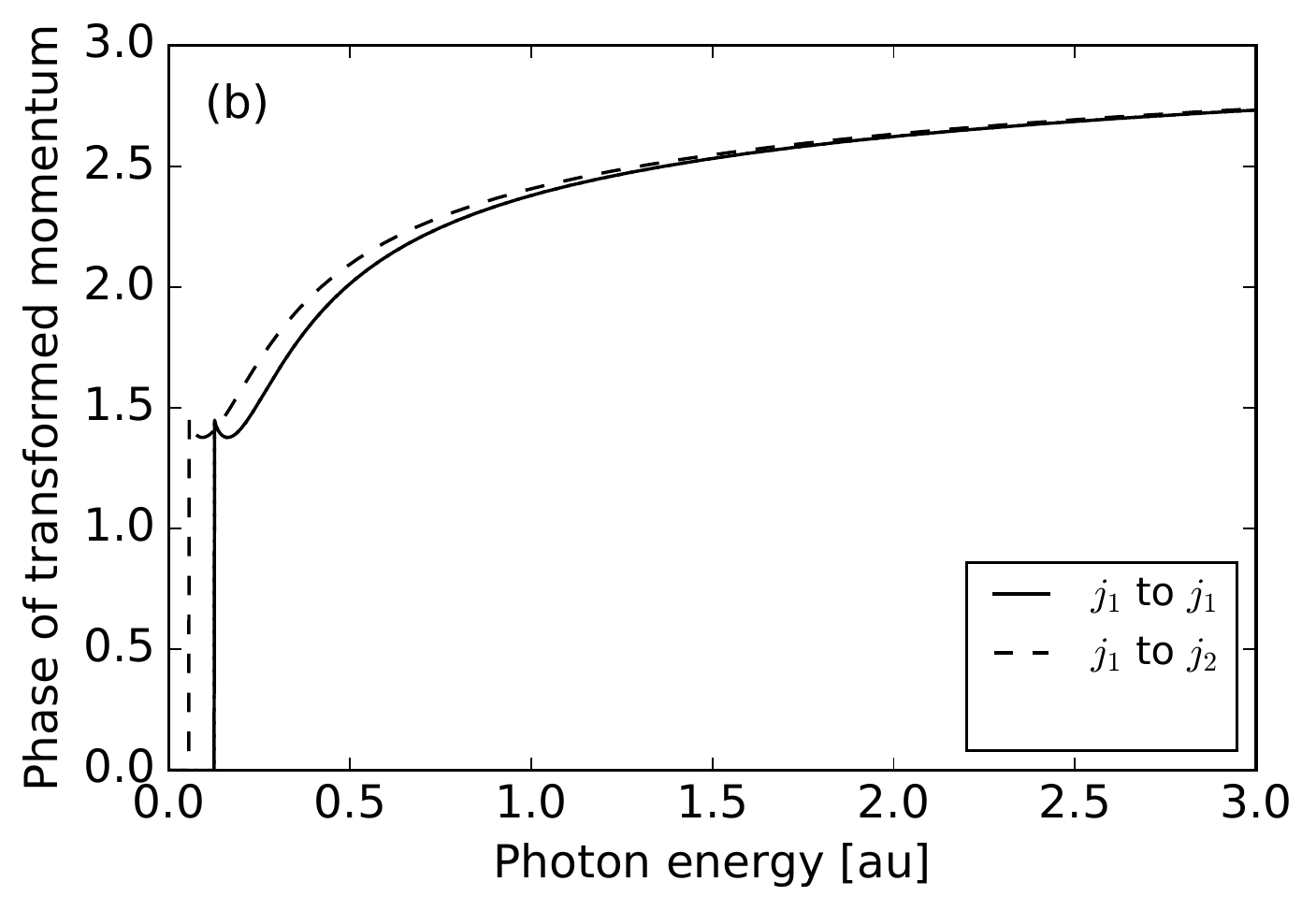}
\caption{(a) The real and imaginary parts of the transformed momentum. The momentum is separated into elementary electron scattering processes according to Eq.~(\ref{pzjj}) with elastic process: $j_1\rightarrow j_1$ and inelastic processes:  $j_1\rightarrow j_2$ and $j_2\rightarrow j_1$. The transformed momentum is normalized to the maximal absolute value of all scattering events. (b) The corresponding phase of the transformed separated momentum. All results are computed using the toy model defined in Eq.~(\ref{toymodel}). 
\label{fig:toymodel}}
\end{figure}
In Fig.~\ref{fig:toymodel}~(a) we show separate components of the transformed momentum, 
\begin{align}
\tilde p_z(\omega)=\sum_{j,j'}\tilde p_z(\omega;j',j), 
\label{pzjj}
\end{align}
corresponding to different electron scattering processes from the initial wave packet state, $\ket{j}$, via all virtual states, $\ket{f}$, to the final wave packet state, $\ket{j'}$. The momentum peaks of the elastic scattering, $j=j'$, are not centered exactly at the central photon energy, $\omega_0=1.5$, because of the decreasing transition amplitude with increasing photon energy in the toy model. The transformed momentum of inelastic scattering, $j\ne j'$, is shifted from the elastic case to higher (lower) frequencies when the electron moves to lower (higher) energy wave packet state. The resonant and principal value contributions to the transformed momentum are of comparable magnitude with a phase of $\sim 2.5$ radians at the central frequency, as shown in Fig.~\ref{fig:toymodel}~(b). 

Eqs.~(\ref{dDEdw}) and (\ref{pzw}) are now combined to find that the absorbed field energy by the atom at $\omega\ge0$ is
\begin{align}
\frac{d\Delta E}
{d\omega} \approx &
-2\omega 
\mathrm{Im}
\left\{
\sum_{ j,j'}
\tilde p_z^-(\omega;j',j) 
\tilde A^*(\omega)
\right\},
\label{dDEdwPT}
\end{align}
where the energy-domain vector potential is 
\begin{align}
\tilde A(\omega) = &
\frac{1}{\sqrt{2\pi}} \int_{-\infty}^{+\infty}dt'
A(t')\exp[i\omega t'] \nonumber \\ \approx & 
e^{i\phi+i\omega t_0}
\frac{A_0}{2\sqrt{2a}}
\exp\left[-\frac{(\omega_0+\omega)^2}{4a}\right], \, \omega<0.
\label{FTA}
\end{align}
In writing Eq.~(\ref{dDEdwPT}) we have also used the RWA,  which is well justified for the parameters in Eq.~(\ref{paperparam}).
While $p_z(\omega)$ [Eq.~(\ref{pzw})] has multiple Gaussian field peaks located at 
$\omega\approx\omega_0\pm\omega_{jj'}$, 
energy can only be absorbed when these peaks are within the bandwidth of the test pulse, $\omega\approx\omega_0$, 
according to Eq.~(\ref{dDEdwPT}). Clearly, no energy can be absorbed at energies where there is no spectral energy density of the test pulse to begin with.       

\begin{figure}[!htb]
\includegraphics[width=0.5\textwidth]{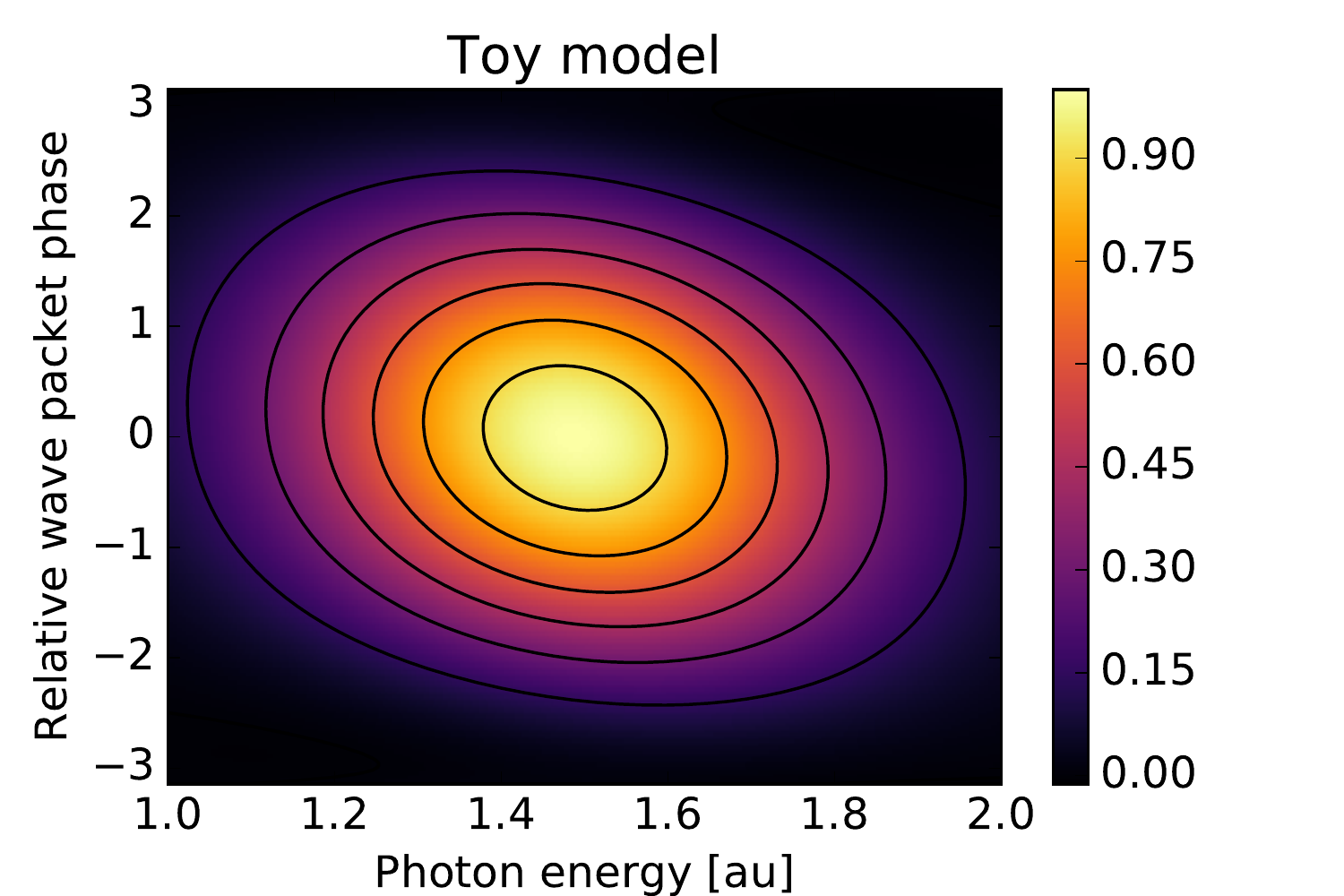}
\caption{The normalized absorbed energy of the toy model for a two-state  wave packet ionized by the test pulse as a function of photon energy, $\omega$, and relative wave packet phase, $\varphi$.   
\label{fig:toymap}}
\end{figure}

As already mentioned, an electron in an initial wave packet state $\ket{j}$ can either transition back to the original wave packet state ($j=j'$) so that $\omega_{jj'}=0$, or it can transition to a different wave packet state ($j\ne j'$) where, in general, $\omega_{jj'}\ne 0$.  
In this way is it possible for the atom to absorb the same photon energy from the test pulse by different electron mechanisms, $\tilde p^-_{z}(\omega;j',j)$, with distinct amplitudes and phases, as shown in Eq.~(\ref{dDEdwPT}). This opens up for interference effects in the absorption of the light provided that the bandwidth of the test pulse is comparable or larger than the wave packet energy splittings of interest, 
\begin{align}
\Delta\omega_\mathrm{e}=
4\ln(2)/\tau_\mathrm{e}>\omega_{jj'}.
\end{align}
We now reintroduce the phase shift between the  $j_2$ and $j_1$ states, $\varphi = \arg[c_{j_2}/c_{j_1}]$, to study how the absorption process can be controlled by a time delayed test pulse. Clearly, the elastic absorption processes  are unaffected by this phase shift since the final wave packet amplitudes is complex conjugated, while the initial is not conjugated, in Eq.~(\ref{pzw}). The inelastic scattering terms, however, are affected by the phase shift. The $j_1\rightarrow j_2$ process suffers a phase reduction of $-\varphi$, while $j_2\rightarrow j_1$ gains a phase of $+\varphi$. Taken together the two inelastic processes leads to a sinusoidal modulation over $\varphi$, while the elastic processes give a constant background in the absorption. In this way, the inelastic processes can increase or decrease the absorption of the test pulse depending on the relative phase. The absorption peak of the toy model is shown in Fig.~\ref{fig:toymap}, as a function of photon energy and relative wave packet phase, using false color.  
The similarities between the exact hydrogen absorption in Fig.~\ref{fig:Hmap} and that of the toy model is striking: The maximal absorption occurs for $\varphi\approx 0$ with a small negative tilt of the peak with increasing photon energy.  

In order to make a quantitative analysis of the phase shifts that maximize the absorption, we Fourier transform the data in Fig.~\ref{fig:toymap} over relative phase, $\varphi$, and directly extract the phase of the oscillation.  
\begin{figure}[!htb]
\includegraphics[width=0.5\textwidth]{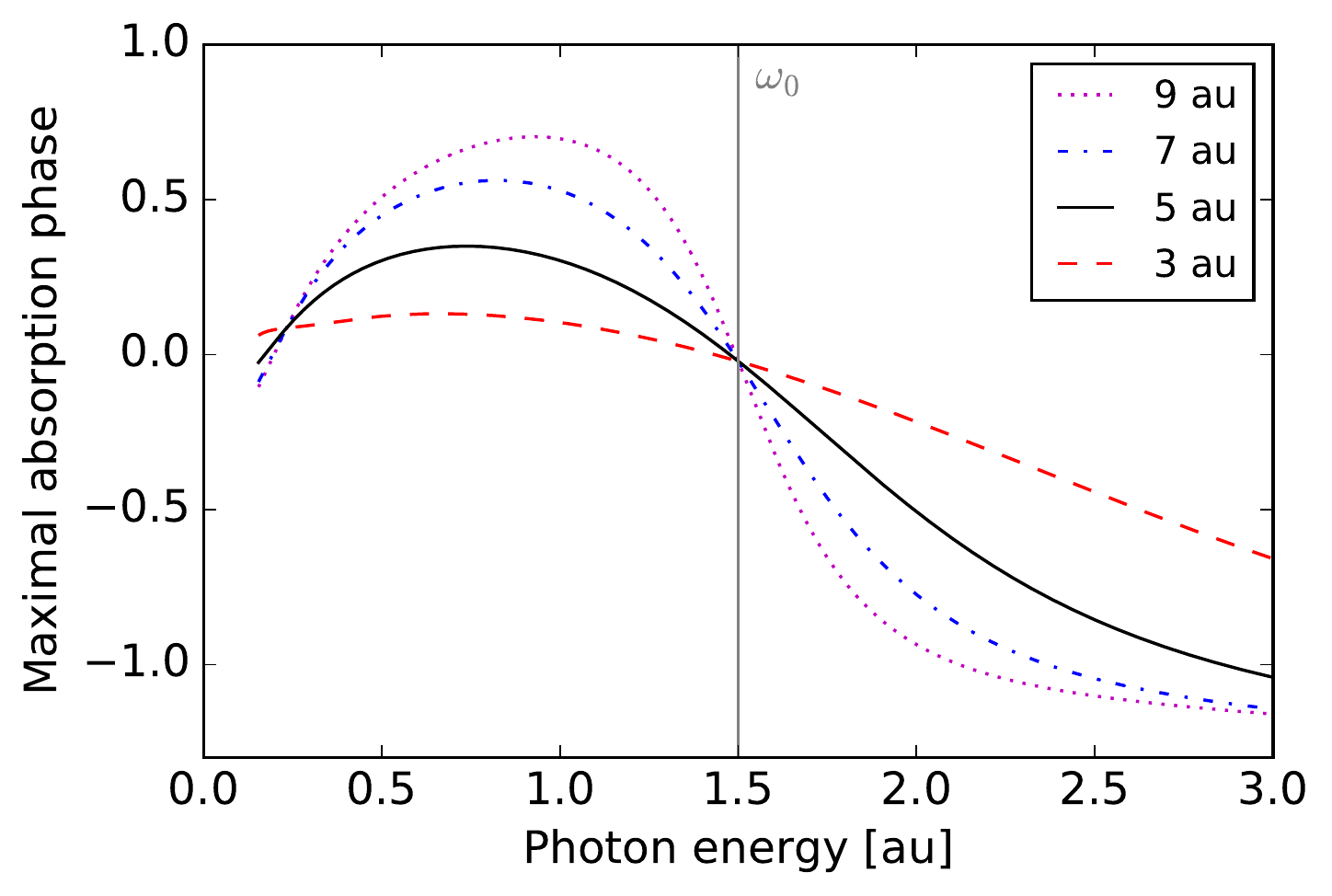}
\caption{Relative phase that maximizes the absorption in the toy model. The maximal absorption phases associated with different pulse envelope durations, $\tau_e=[3,5,7,9]$, are shown for a fixed central photon energy of $\omega_0=1.5$. 
\label{fig:toytau}}
\end{figure}
The phase that maximizes absorption is shown in Fig.~\ref{fig:toytau}. It is seen that the phase exhibits a linear region with a negative slope close the the central frequency of the test pulse. For comparison we also show the maximal absorption phase for a shorter test pulses with $\tau_e=3$ and two longer pulses with $\tau_e=\{7,9\}$. The negative phase slope is clearly dependent on the pulse duration. It appears that a phase shift of $-\pi/2$ is observed across the peak of the field. 

Further away from the central frequency the phase of the toy model exhibits a non-monotonous behavior and close to threshold the phase bends to a positive slope. The exact shape of the phase curve that maximizes absorption depends on the choice of toy model. In the next section we compare how well our toy model, Eq.~(\ref{toymodel}), agrees with the exact result in hydrogen, presented in Fig.~\ref{fig:Hmap}.

\section{Results}
\label{sec:results}
We have found that the absorption of a coherent XUV attosecond test pulse by a bound $2s/3s$ wave packet in hydrogen depends critically on the relative phase of the wave packet. The overall absorption is the largest when the relative phase of the electron wave packet is close to zero. This is not surprising because the test pulse allows for constructive photoionization from both initial wave packet states into the continuum. Conversely, a wave packet with a relative $\pi$-phase ejects photoelectrons that interfere destructively, thus, making the atom more transparent to the test pulse. 
\begin{figure}[!htb]
\includegraphics[width=0.5\textwidth]{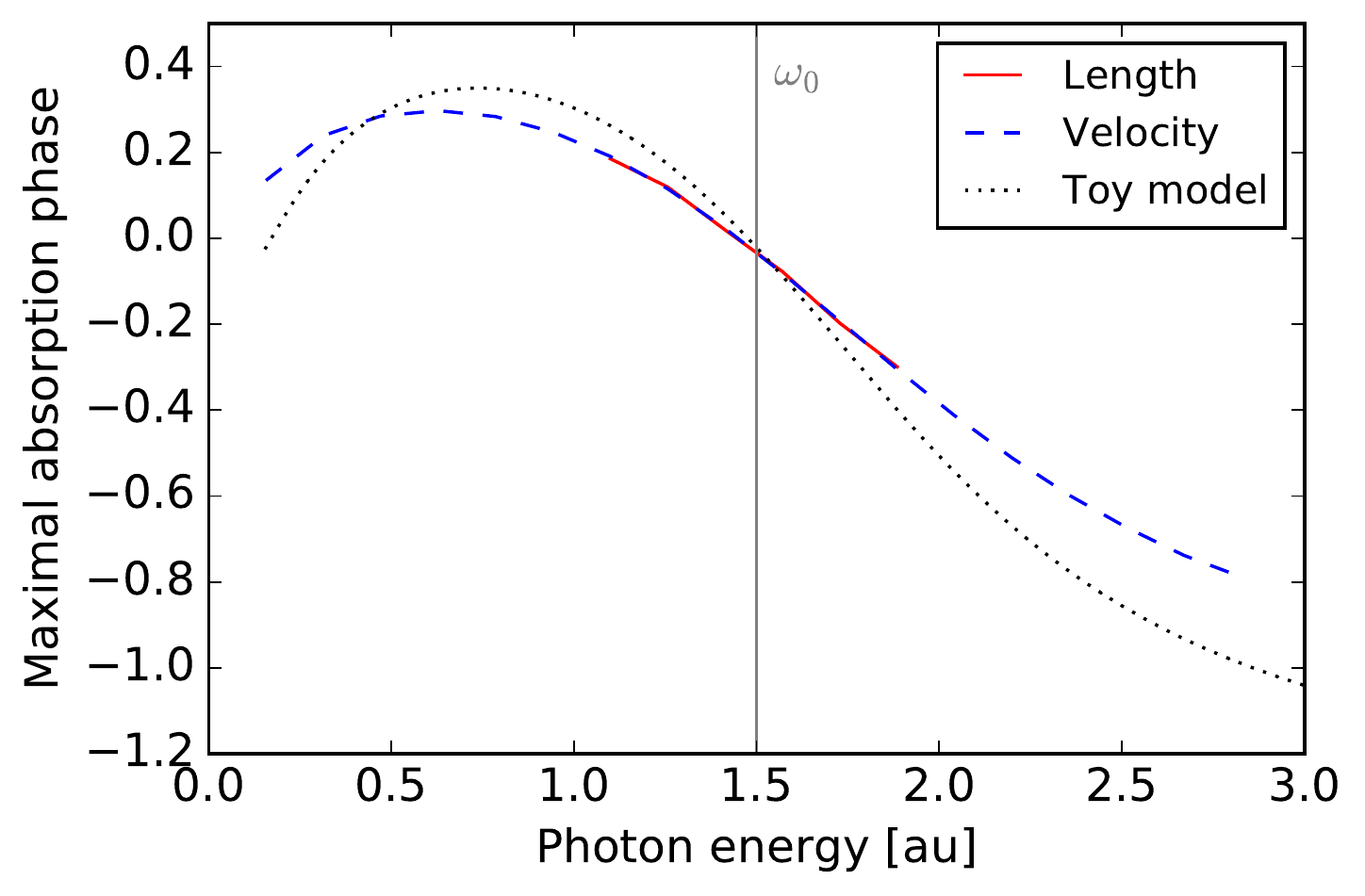}
\caption{Comparison of the $2s/3s$ wave packet relative phase that maximizes the absorption as a function of photon energy for exact case hydrogen and the toy model.    
\label{fig:compareHtoy}}
\end{figure}
In more detail, we have found that the maximal absorption at different photon energies, within the test pulse, occurs at slightly different relative phases. While the central frequency is absorbed the most at zero relative phase, the general curve undergoes a non-monotonous behavior shown in Fig.~\ref{fig:compareHtoy}. 
The numerical hydrogen result shows excellent agreement between the length and velocity gauge. The result is also stable against intensity variations, we have obtained the same curves with a test pulse of 100 times as much intensity, i.e. by changing   from $A_0=0.01$ to $A_0=0.1$.  

In Fig.~\ref{fig:compareHtoy} we also show a good qualitative agreement between the hydrogen calculation and the simple toy model presented in Sec.~\ref{sec:toymodel}. %This shows that the physics to describe the maximal absorption phase originates from virtual excitations of the electron to the continuum with inelastic photon scattering back to the other bound wave packet state. 
The curve is sensitive the model used to describe the dipole interaction. In this sense, the ATAS technique can be used to probe the time-dependent photoelectron wave packet, Eq.~(\ref{cf}), by inelastic photon scattering back onto the original bound wave packet. 
%, but we also have showed that the phase depends on the pulse shape.  

%
\begin{figure}[!htb]
\includegraphics[width=0.5\textwidth]{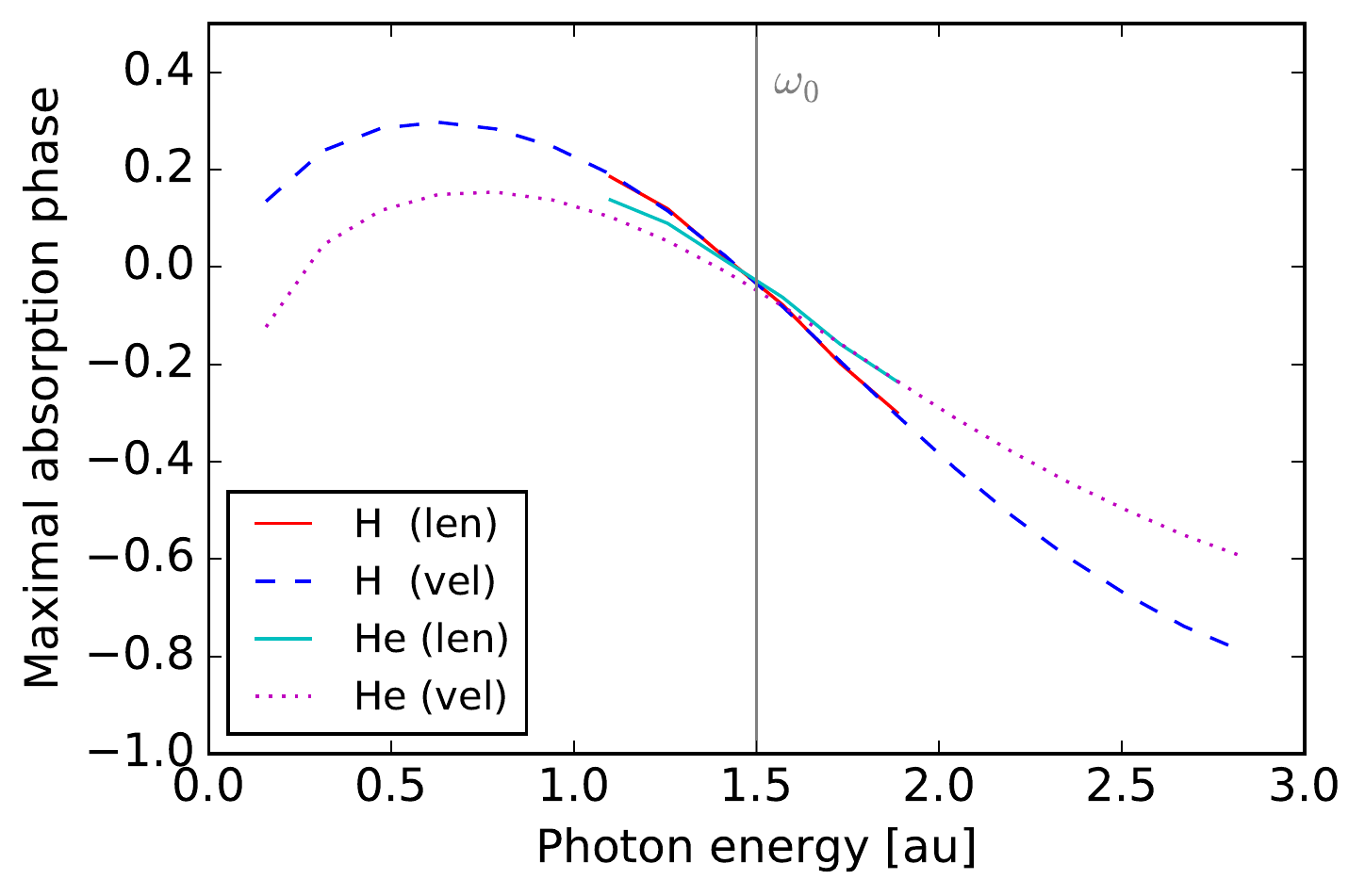}
\caption{Comparison of the $2s/3s$ wave packet relative phase that maximizes the absorption as a function of photon energy for exact case hydrogen and helium.
\label{fig:compareHHe}}
\end{figure}
To investigate the generality of our result we have performed calculations for helium within TDCIS for an initial $1s^{-1}2s/3s$ wave packet. As expected the absorption is qualitatively the same. One difference is that we do not obtain gauge invariance, but is a known issue with TDCIS and not specific to our method. The velocity gauge calculation is converged best and we only show the converged length gauge result close to the central frequency in Fig.~\ref{fig:compareHHe}. Due to the nature of how the helium calculations were made we can not assign an exact phase between the two initial bound states. We have shifted the relative phase to coincide with the crossing at the central photon energy found in hydrogen. 

\section{Conclusions}
\label{sec:conclusions}
In conclusion we have found that absorption of an attosecond pulse can be controlled by the relative phase of a prepared atom wave packet. The ATAS method can be used to probe virtual ionization processes, where a photoelectron is first ejected into the continuum, but then may (in)elastically scatter back to the initial bound wave packet. Surprisingly we find that the absorption is not exactly maximal for zero relative phase of the wave packet, but instead a negative phase variation is found that undergoes a $-\pi/2$ shift over the bandwidth of the test pulse. 

Recently a similar phase effect was observed theoretically in strong photoionization of a bound wave packet by a strong IR Gaussian pulse, where the phase effect was attributed to the  time- and intensity-dependent depletion of the initial wave packet states \cite{Pazourek2016}. In our case, however, the phase effect is not due to depletion of the initial wave packet states, but rather due to the time-dependent evolution of the photoelectron wave packet. Our phase effect occurs already at extremely weak field strengths and it is found to be stable against an increased test pulse intensity. This intrinsic phase lag of the ATAS method is problematic for accurate characterization of the test pulse because it makes it difficult to disentangle which part of phase contribution comes from the stimulated atomic transitions and which come from the spectral phases of the test pulse. Since the phase effect in ATAS is quite large it can not be assumed to be small compared to the group delay of the attosecond test pulse in general.   

\begin{acknowledgments} 
We thank Andreas Wacker for discussions. 
J.M.D. is funded by the Swedish Research Council, Grant No. 2014-3724. 
S.P. is funded by the the NSF through a grant to ITAMP.
E.L. is funded by the Swedish Research Council, Grant No. 2016-03789.

\end{acknowledgments}

%=======================================================================
%

% REFERENCES


\end{document}